\documentclass[%
 reprint,
 amsmath,amssymb,
 aps,
 pra, 
 longbibliography
]{revtex4-2}

\usepackage{footmisc,multirow}
\usepackage{subfigure} 
\usepackage{amsmath,slashed}
\usepackage{upgreek}
\usepackage{graphicx}
\begin{document}

\preprint{IPPP/23/50}

\title{
Deuterium spectroscopy for enhanced bounds on physics beyond the Standard Model}

\author{Robert M. Potvliege}
\email{r.m.potvliege@durham.ac.uk}
\affiliation{Department of Physics, Durham University, South Road, Durham DH1 3LE, United Kingdom}

\author{Adair Nicolson} 
\altaffiliation{Present address: UCL Department of Chemistry, University College London, 20 Gordon Street, Kings Cross, London WC1E 6BT, United Kingdom}

\author{Matthew P. A. Jones}  
\email{m.p.a.jones@durham.ac.uk}
\affiliation{Department of Physics, Durham University, South Road, Durham DH1 3LE, United Kingdom}

\author{Michael Spannowsky}
\email{michael.spannowsky@durham.ac.uk}
\affiliation{Department of Physics, Institute of Particle Physics Phenomenology, Durham University, South Road, Durham DH1 3LE, United Kingdom}


\begin{abstract}
We consider the impact of combining precision spectroscopic measurements made in atomic hydrogen with similar measurements made in atomic deuterium on the search for physics beyond the Standard Model. Specifically we consider the wide class of models that can be described by an effective Yukawa-type interaction between the nucleus and the electron. We find that  it is possible to set bounds on new light-mass bosons that are orders of magnitude more sensitive than those set using a single isotope only, provided the interaction couples differently to the deuteron and proton. Further enhancements of these bounds by an order of magnitude or more would be made possible by extending the current measurements of the isotope shift of the 1s$_{1/2}$~--~2s$_{1/2}$ transition frequency to that of a transition between the 2s$_{1/2}$ state and a Rydberg s-state.
\end{abstract}

\maketitle

\section{Introduction}
\label{sec:intro}

The exquisite precision now achievable in optical frequency measurements and the ongoing development of methods for cooling and trapping atoms, molecules, and highly charged ions suscitate a growing interest in searching for new physics beyond the Standard Model using precision spectroscopy \cite{Jaeckel2010,Karshenboim2010,Karshenboim2010b,Brax2011,Brax2015,Delaunay2017,Jones2020,Delaunay2023}.
Spectroscopy experiments already complement existing measurement strategies from high-energy physics experiments and astrophysical observations \cite{Safronova2018}.

However, one of the biggest challenges to fully exploiting the measurement precision currently achievable is the difficulty of direct comparison with the predictions of the Standard Model for many-electron atoms. The achievable precision in calculations of transition frequencies for these systems is limited by the difficulty of exactly solving the many-electron Schr\"odinger equation to the required level of accuracy and, more fundamentally, by a lack of knowledge about the necessary many-electron quantum electrodynamics (QED) corrections.

Two approaches round this problem have been proposed. 
The first is based on many-electron atoms but reduces the need for precise electronic structure calculations by considering different isotopes of the same species \cite{Delaunay2017b,Berengut2018,Ohayon2019}. Since it was first proposed this method has been applied to experiments in trapped ions \cite{Counts2020,Solaro2020,Hur2022,Rehbehn2023}. However, the identification of Physics beyond the Standard Model in this approach is complicated by computational difficulties \cite{Viatkina2023} and by the need of using sufficiently accurate models of the nuclei considered \cite{Allehabi2021,Muller2021,Munro-Laylim2022}. 

The other is to use hydrogenic atoms,
for which the necessary QED calculations can often be done to a precision matching the experimental error on the measured transitions. Early work in this direction extended the isotope shift method to isotopes of hydrogen and helium \cite{Delaunay2017}. Subsequently our group showed that direct-experiment theory comparison across an entire set of spectroscopic data (in this case for hydrogen--1) could be used to set global bounds on beyond Standard Model forces \cite{Jones2020}. This approach was recently significantly extended by Delaunay et al., who performed a global fit to the entire set of relevant CODATA measurements (not just hydrogen-like atoms) \cite{Delaunay2023}.

In this paper we use both the isotope shift and global constraint approaches to set new bounds using spectroscopic measurements in the electronic and muonic isotopes of hydrogen only. Measurements in ordinary hydrogen set a bound on the product of the constants $g_e$ and $g_p$ parametrizing, respectively, how the electron and the proton couple to a hypothetical New Physics (NP) boson. The presence of a neutron in the deuterium nucleus introduces a different product of coupling constants for this species, i.e., $g_e g_d$ rather than $g_e g_p$. It is reasonable to assume that $g_d$ differs significantly from $g_p$. For example, it has been noted that the beryllium anomaly \cite{Krasznahorkay2016, Krasznahorkay2018}
can be explained by a light NP boson with a mass of ${\sim}17$~MeV that couples potentially very differently to protons and neutrons \cite{Feng2017,Barducci2022}. We find that the overall bound on possible fifth forces is extremely sensitive to the ratio $g_d/g_p$ with the global bound for both isotopes rapidly exceeding that set by hydrogen--1 alone if $g_d/g_p \neq 1$. E.g. for both $g_d/g_p=2$ and $g_d/g_p = 0$, we find that the upper bound on the possible value of $|g_eg_p|$ for NP bosons in the mass range $1-10$ eV is strengthened by two orders of magnitude compared to that set using hydrogen--1 data alone, and by a further factor of five when the Lamb shift measurements in muonic hydrogen and muonic deuterium are also taken into account.

The theoretical model of NP interaction considered in this work is outlined in Section~\ref{sec:param}. Compared to previous work, we relax a number of assumptions: we do not make any assumptions about the New Physics model being tested beyond those required to get to the Yukawa potential, and we do not constrain the ratio of the coupling to the deuteron and the proton, or the sign of the NP shift. Bounds based on the whole of the current high-precision spectroscopic data are presented in Section~\ref{sec:results}. Bounds based only on the isotope shift of the 1s$_{1/2}$~--~2s$_{1/2}$ interval \cite{Delaunay2017} are also presented in this section.
Prospects for further tightening the latter are discussed in Section~\ref{sec:isotopeshift}. A recap of our main results is given in the Conclusions, Section~\ref{sec:conclusions}. The main body of the paper is complemented by five appendices devoted to more technical details, including a discussion of the impact of an NP interaction on the determination of the proton and deuteron charge radii from Lamb shift measurements in the muonic species. 

\section{New Physics Scenarios}
\label{sec:param}
Most extensions of the Standard Model, which aim to explain the observation of excesses in the kinematic distributions of collision events or decays, require the introduction of propagating degrees of freedom that manifest as particles. To test the existence of such degrees of freedom and their interactions with Standard Model particles, it has become popular to parametrise a deformation of the Standard Model Lagrangian in terms of so-called simplified models \cite{Alves2012}.

Assuming a new force to be mediated through a spin-0 particle $X_0$ that couples to leptons and quarks with couplings $g_{l_i}$ and $g_{q_i}$ respectively, we can expand the Standard Model Lagrangian $\mathcal{L_\mathrm{SM}}$  by 
	\begin{equation}
	\mathcal{L} = \mathcal{L}_{\rm SM} +
	\sum_{i} \left[g_{l_{i}} \bar{l}_i   l_i + g_{q_{i}} \bar{q}_i   q_i \right ] X_0.
\label{eq:intnew}
\end{equation}
Here $i$ denotes the three flavor generations, and $l_i$ and $q_i$ refer to the mass basis of the SM fermions~\cite{noteaboutmodels}

With the Lagrangian of Eq.~(\ref{eq:intnew}), the interaction mediated by the
NP boson $X_0$ between an atomic nucleus and an electron or a muon contributes an additional Yukawa potential $V_{\rm NP}(r)$ to the Hamiltonian.
In natural units,
\begin{equation}
V_{\rm NP}(r) = -\frac{g_l g_N}{4 \pi}~\frac{1}{r}~e^{-m_{X_0} r},
\label{eq:yukpot1}
\end{equation}
where $r$ is the distance between the electron or muon and the nucleus and $m_{X_0}$ is the particle's mass. Higher integer-spin mediators, e.g., vector particles, would also give rise to a Yukawa potential of this form. In general, for a massive force mediator of integer spin $s$, 
\begin{equation}
V_{\rm NP}(r) = (-1)^{s+1}\frac{g_l g_N}{4 \pi}~\frac{1}{r}~e^{-m_{X_0} r},
\label{eq:yukpot2}
\end{equation}
However, assuming Lorentz invariance and the unitarity of the transition matrix element leads to an attractive (repulsive) force if $g_l g_N > 0$ ($g_l g_N < 0$) in the case of an even-spin mediator and to an attractive (repulsive) force if $g_l g_N < 0$ ($g_l g_N > 0$) in the case
of an odd-spin mediator. To remain general, we allow positive and negative values for the product $g_l g_N$.
We denote the coupling constant $g_l$ by $g_e$ for the case of an electron and $g_\mu$ for the case of a muon, and the coupling constant $g_N$ by $g_p$ for the case of a proton and $g_d$ for the case of a deuteron.

Such an NP interaction would shift the energy of an $n,l$ state of hydrogen or deuterium by a quantity $\delta E^{\rm NP}_{nl}$. The interaction is expected to be very weak. This energy shift does not need to be calculated beyond the first order in $V_{\rm NP}(r)$. The case of electronic hydrogen and electronic deuterium is straightforward:
as noted, e.g., in our previous publication on this topic \cite{Jones2020}, referred to as paper I in the following,
\begin{equation}
    \delta E^{\rm NP}_{nl} =
    \int_0^\infty \big|R^{}_{nl}(r)\big|^2\, \frac{B\exp(-Cr)}{r}\,
    r^2\,{\rm d}r,
\label{eq:shift1}
\end{equation}
where $R_{nl}(r)$ is the non-relativistic radial wave function of the unperturbed state. The interaction potential is written in terms of the constants $B$ and $C$ rather than $g_lg_N$ and $m_{X_0}$ to facilitate conversion from natural units to atomic units. To five significant figures,
\begin{equation}
    B\, [E_{\rm h}\,a_0] = 10.905\, (-1)^{s+1}\,g_lg_N,
    \label{eq:Bhydrogen1}
\end{equation}
and
\begin{equation}
    C\, [a_0^{-1}] = 2.6817 \times 10^{-4}\, m_{X_0}\,[\mbox{eV}].
\end{equation}
As defined by the above equations, the coupling constant $g_lg_N$ is a pure number.

Relativistic wave functions must be used for muonic hydrogen and muonic deuterium. This case is addressed in Appendix~\ref{sec:rprd}.

\section{Current bounds on the strength of an NP~interaction}
\label{sec:results}

\subsection{Methods}
\label{sec:methods}

We only consider spectroscopic data in the present work, in view of the difficulty of deriving unambiguous results from the existing high-precision scattering data \cite{Khabarova2021}. 

For electronic hydrogen (eH) and electronic deuterium (eD), the relevant spectroscopic data consists of a set of transition frequencies measured to a high degree of precision, i.e., $\nu_{b_1a_1}^{\rm exp}$, 
$\nu_{b_2a_2}^{\rm exp}$, 
$\nu_{b_3a_3}^{\rm exp}$, etc. As in paper I, we obtain bounds on the possible strength of a hypothetical NP interaction by comparing these measured frequencies to the prediction of the theoretical model outlined in Section~\ref{sec:param}. 
Namely, we compare each measured transition frequency $\nu_{b_ia_i}^{\rm exp}$ to its theoretical counterpart, $\nu_{b_ia_i}^{\rm th}$, the latter being given by its Standard Model value corrected for the NP shift defined by Eq.~(\ref{eq:shift1}):
\begin{equation}
\nu_{b_ia_i}^{\rm th} =
    \nu_{b_ia_i}^{\rm SM} + \nu_{b_ia_i}^{\rm NP}, \quad i = 1,2,3,\ldots
\end{equation}
For a transition between a state $a$ of principal quantum number $n_a$ and orbital angular momentum quantum number $l_a$ and a state $b$ of principal quantum number $n_b$ and orbital angular momentum quantum number $l_b$,
\begin{equation}
    \nu_{ba}^{\rm NP} = (\delta E_{n_bl_b}^{\rm NP} - \delta E_{n_al_a}^{\rm NP})/h,
\label{eq:DeltaNPdefined}
\end{equation}
where $h$ is Planck's constant. Each SM transition frequency $\nu_{b_ia_i}^{\rm SM}$ is the sum of a gross structure contribution, $\nu_{b_ia_i}^{\rm g}$, and of various relativistic, QED and hyperfine corrections \cite{Mohr2016,Yerokhin2019,Tiesinga2021}. These various contributions depend on the Rydberg frequency (${\cal R}$), and some also depend on the charge radius of the proton ($r_p$) and the charge radius of the deuteron ($r_d$). As in Paper~I, we group them into a gross structure term depending sensitively on ${\cal R}$, terms depending sensitively on $r_p$ or $r_d$, and a term accounting for all the other contributions to the SM transition frequency. Specifically, we write 
\begin{equation}
    \nu_{b_ia_i}^{\rm SM} = {\cal R}\,\tilde{\nu}_{b_ia_i}^{\rm g}
    + r_p^2\,\tilde{\nu}_{b_ia_i}^{\rm ps} + r_d^2\,\tilde{\nu}_{b_ia_i}^{\rm ds} + \nu_{b_ia_i}^{\rm oc},
    \label{eq:Deltath}
\end{equation}
where the factors $\tilde{\nu}_{b_ia_i}^{\rm g}$, $\tilde{\nu}_{b_ia_i}^{\rm ps}$ and $\tilde{\nu}_{b_ia_i}^{\rm ds}$ and the term ${\nu}_{b_ia_i}^{\rm oc}$ do not strongly depend on the precise values of ${\cal R}$, $r_p$ and $r_d$, if they depend on them at all. In particular, 
\begin{equation}
    \tilde{\nu}_{b_ia_i}^{\rm g} = \nu_{b_ia_i}^{\rm g}/{\cal R} =\left(\frac{1}{n_{a_i}^2}-\frac{1}{n_{b_i}^2}\right)\,
    \frac{m_{\rm r}}{m_e},
    \label{eq:gross}
\end{equation}
where $m_{\rm r}$ the reduced mass of the atom and $m_{e}$ the mass of the electron. The proton size and deuteron size terms,  $r_p^2\,\tilde{\nu}_{b_ia_i}^{\rm ps}$
and $r_d^2\,\tilde{\nu}_{b_ia_i}^{\rm ds}$, group the QED contributions which are roughly proportional to, respectively, $r_p^2$ and $r_d^2$, and the
term ${\nu}_{b_ia_i}^{\rm oc}$ encapsulates all the other corrections to the gross structure contribution.

The data for muonic hydrogen ($\mu$H) and muonic deuterium ($\mu$D) are currently limited to high precision measurements of the Lamb shift of the $n=2$ states. In principle, this data could be integrated in the calculation of NP bounds exactly as described in the previous paragraph. However, it is more convenient to use a slightly different (albeit equivalent) formulation, which is more closely related to the QED theory of these species.  Namely, we write the theoretical Lamb shift of muonic hydrogen ($\Delta E^{\rm th}_{\mu{\rm H}}$) 
as a sum of a term proportional to the square of the proton charge radius, of a term which does not depend sensitively on this radius, and of a New Physics contribution, and similarly for the theoretical Lamb shift of muonic deuterium ($\Delta E^{\rm th}_{\mu{\rm D}}$): 
\begin{align}
    \Delta E^{\rm th}_{\mu{\rm H}} &= \Delta E^{\rm main}_{\mu{\rm H}} +
r_p^2
    \tilde{\Delta}E^{\rm ns}_{\mu{\rm H}} 
    + \Delta E^{\rm NP}_{\mu{\rm H}},\\
    \Delta E^{\rm th}_{\mu{\rm D}} &= \Delta E^{\rm main}_{\mu{\rm D}} + 
r_d^2
    \tilde{\Delta}E^{\rm ns}_{\mu{\rm D}} 
    + \Delta E^{\rm NP}_{\mu{\rm D}}.
\end{align}
The QED theory of these species \cite{Antognini2013,Lensky2022} yields $\Delta E^{\rm main}_{\mu{\rm H}} = 206.0668(25)$~meV,  $\tilde{\Delta}E^{\rm ns}_{\mu{\rm H}} = -5.2275(10)$~meV/fm$^2$,
$\Delta E^{\rm main}_{\mu{\rm D}} = 230.5283(200)$~meV and  $\tilde{\Delta}E^{\rm ns}_{\mu{\rm D}} = -6.10801(28)$~meV/fm$^2$. We calculate the New Physics correction terms $\Delta E^{\rm NP}_{\mu{\rm H}}$ and $\Delta E^{\rm NP}_{\mu{\rm D}}$ as described in Appendix~\ref{sec:rprd} (the calculation is based on the Dirac equation and takes into account the nuclear charge distribution and vacuum polarization). Equating $\Delta E^{\rm th}_{\mu{\rm H}}$ and $\Delta E^{\rm th}_{\mu{\rm D}}$ to the experimental Lamb shifts \cite{Antognini2013,Pohl2016} yields NP-corrected values for $r_p$ and $r_d$. 

We co-determine ${\cal R}$, $r_p$ and $r_d$ and confidence levels of the NP coupling constants by a global correlated $\chi^2$-fit of our theoretical model to experiment \cite{Mohr2000}. Specifically, given values of $g_p$, $g_d$, $g_e$, $g_\mu$ and $m_{X_0}$ characterizing an NP interaction, we adjust ${\cal R}$, $r_p$ and $r_d$ so as to minimize the difference between the theoretical transition frequency $\nu_{b_ia_i}^{\rm th}$ and the corresponding experimental transition frequency $\nu_{b_ia_i}^{\rm exp}$ for each of the transitions considered. In calculations including the muonic species, we also minimize the difference between the charge radii derived from measurements in the electronic species from those derived from measurements in the muonic species.
The resulting value of $\chi^2$ characterizes how close the theory fits the data for given values of $g_p$, $g_d$, $g_e$, $g_\mu$ and $m_{x_0}$. Having this value, we calculate the upper tail cumulative distribution function $Q(\chi^2|\nu) = 1 - P(\chi^2|\nu)$ for the relevant number of degrees of freedom, $\nu$. $Q(\chi^2|\nu)$ is the probability that the difference between theory and experiment arises only from random experimental and theoretical errors. A small value of this probability indicates a low likelihood that the data are compatible with an NP interaction of the type considered in this work. The calculation is essentially the same as that underpinning the determination of ${\cal R}$, $r_p$ and $r_d$ by the Committee on Data of the International Council for Science (CODATA) \cite{Mohr2016,Tiesinga2021}, albeit here limited to the spectroscopic data and generalized to the encompass the possibility of an NP interaction. Further details can be found in paper I and in Appendix~\ref{sec:notes}.

\subsection{Bounds based on spectroscopic measurements in eH and eD}

\begin{figure}[t!]
\centering
$\mbox{}$\\[-2.0cm]
\includegraphics[width=\columnwidth]{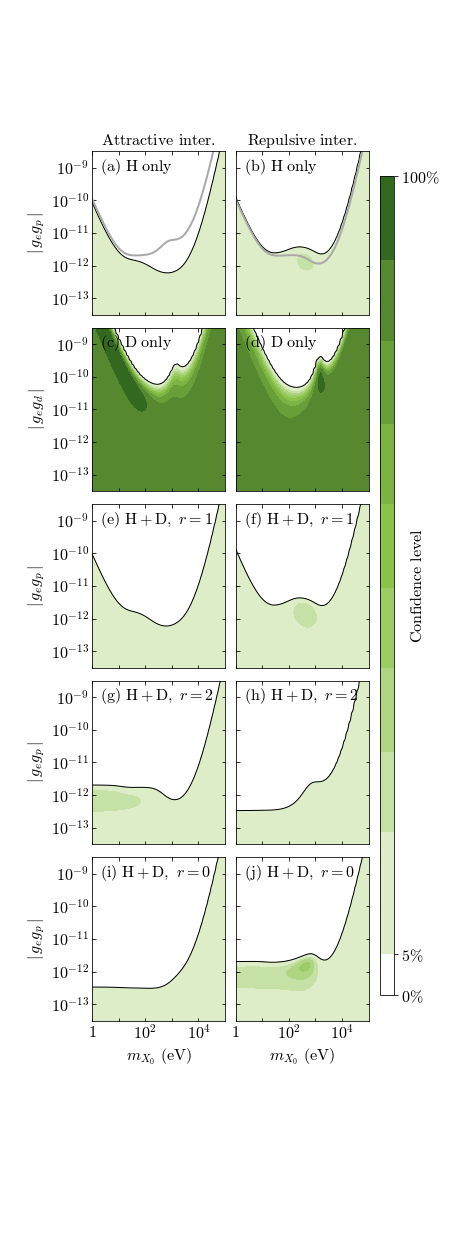}
$\mbox{}$\\[-4.2cm]
\caption{Confidence level that an NP interaction is compatible with the World spectroscopic data for electronic hydrogen-1 and/or electronic deuterium.
(a) and (b): Results solely based on the hydrogen-1 data. (c) and (d): Results solely based on the deuterium data. (e) to (j): Results obtained by combining the hydrogen-1 and deuterium data, assuming that the ratio $r = g_d/g_p$ is either 1, 2 or 0. The boxes on the left side of the figure refer to an attractive NP interaction, the others to a repulsive NP interaction.
The possibility of an NP interaction with parameters falling in a white region is excluded at the 95\% confidence level. Black curves: the corresponding upper bounds on the value of $g_eg_p$ or $g_eg_d$. Grey curves: the upper bounds on the value of $g_eg_p$ found in Ref.~\cite{Jones2020}. 
}
\label{fig:fig1_10boxes1p0}
\end{figure}

Fitting solely against the hydrogen data yields the results shown in Fig.~\ref{fig:fig1_10boxes1p0}(a) for an attractive NP interaction and Fig.~\ref{fig:fig1_10boxes1p0}(b) for a repulsive NP interaction. The shades of green indicate the extent to which the data are compatible with the presence of the interaction, as per the colour axis. The white regions correspond to areas where an NP interaction is excluded at the 95\% confidence level (i.e., areas where the probability that the data is compatible with an NP interaction is less than 0.05). We take the boundaries delineating the white and green regions as defining upper bounds on the values of $g_eg_p$ consistent with the current experimental evidence, as determined in this particular analysis. These bounds are highlighted by black curves for better visibility. For comparison, the grey curves represent the corresponding bounds found in paper~I for the largest data set considered in this previous work, ``data set A''. This data set did not include recent measurements, which are taken into account in the present work (namely,  measurements of the 2p$_{1/2}$~--~2s$_{1/2}$, 1s$_{1/2}$~--~3s$_{1/2}$ and 2s$_{1/2}$~--~8d$_{5/2}$ intervals \cite{Bezginov2019,Grinin2020,Brandt2022}). Compared to paper~I, the bounds found for the more extensive data set considered here are tighter for an attractive interaction and slightly less tight for a repulsive interaction, except in the low mass region where they are similar. However, it should be noted that the confidence levels remain low at all values of $g_eg_p$, reflecting the well-known inconsistencies between some of the data.

Fitting solely against the deuterium data  yields the results shown in
Figs.~\ref{fig:fig1_10boxes1p0}(c) and (d). 
These data exhibit a higher consistency level than hydrogen, which translates into higher confidence levels at low values of $|g_eg_d|$. For mediator masses below 100~eV, the bounds on $|g_eg_d|$ set by the deuterium data are considerably less stringent than those on $|g_eg_p|$ set by the hydrogen data.
The latter is strengthened in this region by measurements on high Rydberg states, which are not available for deuterium \cite{Karshenboim2010,Jones2020}.

Combining the hydrogen data with the deuterium data yields the results shown in Figs.~\ref{fig:fig1_10boxes1p0}(e)--(j), for three different ratios of the respective coupling constants. We denote this ratio by $r$:
\begin{equation}
r = g_d/g_p.
\label{eq:rdefined}
\end{equation}
The confidence levels for $g_d = g_p$ ($r = 1$) hardly differ from those solely based on the hydrogen data, which should be expected since the hydrogen data place a stronger constraint on the product $g_eg_p$ than the deuterium data does on the product $g_eg_d$.

However, taking $g_d \not= g_p$
creates a mismatch in the NP shift of the transition frequencies between the two isotopes, 
which generally diminishes the data's compatibility with the presence of an NP interaction. As shown by these figures, the mismatch created by assuming that $g_d = 2 g_p$ ($r = 2$) or that $g_d = 0$ ($r =0$) rather than
$g_d = g_p$ does not significantly impact the bound on $g_eg_p$ for masses above 1~keV. For lower masses, however, it shifts this bound towards substantially smaller values of $|g_eg_p|$.

\begin{figure}[t!]
\centering
\includegraphics[width=\columnwidth]{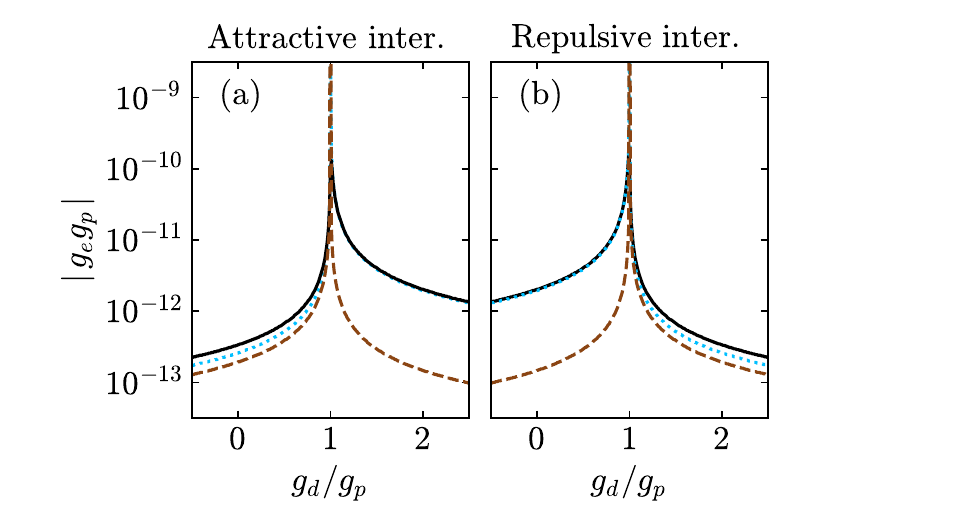}
\caption{The value of $|g_eg_p|$ above which an NP interaction is excluded at the 95\% confidence level, assuming that $m_{X_0} = 1$~eV, vs.\ the ratio $g_d / g_p$. (a): Attractive NP interaction. (b): Repulsive NP interaction. Solid black curves: results based on the World spectroscopic data, as in Figs.~\ref{fig:fig1_10boxes1p0}(e) to (j). Blue dotted curves: the same but now excluding the results for the high Rydberg states. Dashed brown curves: isotope shift results as in Fig.~\ref{fig:fig1_4boxes}.
}
\label{fig:variable_ratio_2023}
\end{figure}
As illustrated by Fig.~\ref{fig:variable_ratio_2023} this bound on $g_eg_p$ tightens very rapidly in the low mass region as soon as $r$ starts departing from 1. Fig.~\ref{fig:variable_ratio_2023} also shows that it is only for $r=1$ that the high Rydberg state datum is important for constraining $g_e g_p$ (compare the dotted blue curves to the solid black curves): for any other values of $r$, the bounds on $g_eg_p$ depend little on whether the measurements on Rydberg states are taken into account.

One may also note that the results of Fig.~\ref{fig:fig1_10boxes1p0}(a)--(j) tend to favor non zero values of $g_eg_p$ in certain ranges of mediator masses; however, the difference in confidence level with $g_eg_p = g_eg_d = 0$ is too small to point towards the possible existence of an NP interaction.

\subsection{Bounds based on spectroscopic measurements in eH, eD, $\mu$H and $\mu$D}
\label{sec:allres}

Adding the muonic hydrogen and muonic deuterium measurements to the data used in the previous section makes it necessary to ascertain the role of an NP interaction on the intervals measured in these species. The issue is discussed in Appendix~\ref{sec:rprd}. We find that an NP interaction with a carrier mass below 10~keV is unlikely to shift these intervals significantly, unless this interaction would couple to a muon much more strongly than to an electron ($|g_\mu| \gg |g_e|$). However, the situation is less clear for higher carrier masses, even for $g_\mu \approx g_e$. 

Accordingly, we allow for the possibility that an NP interaction plays a role in the measurements of the muonic species.
As described in Section~\ref{sec:methods}, we do this by correcting
$r_p$ and $r_d$ for a hypothetical NP shift and use these corrected values when fitting theory to experiment. Bounds can also be obtained by following the method outlined in Appendix~\ref{sec:othermethod}, which is more limited in scope and produces almost identical results where comparison is possible. Either way, the confidence levels derived from these calculations depend on the relative values of the coupling constants $g_\mu$ and $g_e$. For simplicity, we assumed that $g_\mu = g_e$ for producing the figures presented in this section. Results calculated for other relative values of $g_\mu$ can be found in Table~\ref{table:gmutoge} and are discussed below. 

\begin{figure}[t!]
\centering
\includegraphics[width=\columnwidth]{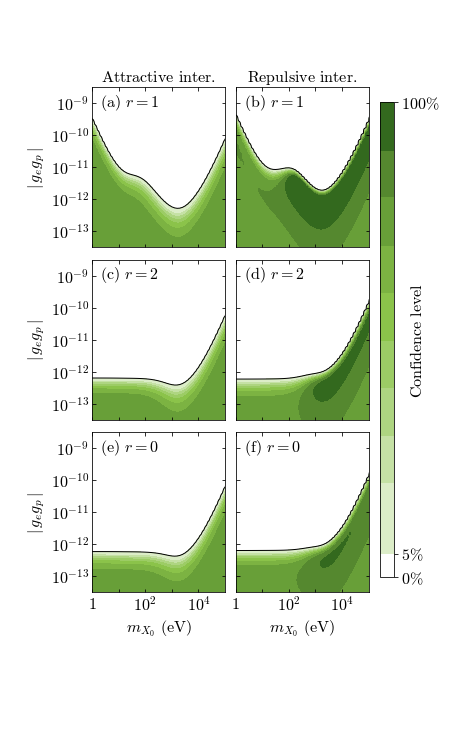}

$\mbox{}$\\[-2.3cm]
\caption{
Confidence level that an NP interaction is compatible with the World spectroscopic data for electronic hydrogen-1, electronic deuterium, muonic hydrogen and muonic deuterium,
assuming that the ratio $r = g_d/g_p$ is either 1, 2 or 0 and that $g_\mu = g_e$. The boxes on the left side of the figure refer to an attractive NP interaction, the others to a repulsive NP interaction.
All experimental uncertainties have been increased by 60\%.
As in Fig.~\ref{fig:fig1_10boxes1p0}, the possibility of an NP interaction with parameters falling in a white region is excluded at the 95\% confidence level and the black curves indicate the corresponding upper bounds on the value of $g_eg_p$.
}
\label{fig:fig1_6boxes1p6}
\end{figure}
\begin{figure}[t!]
\centering
\includegraphics[width=\columnwidth]{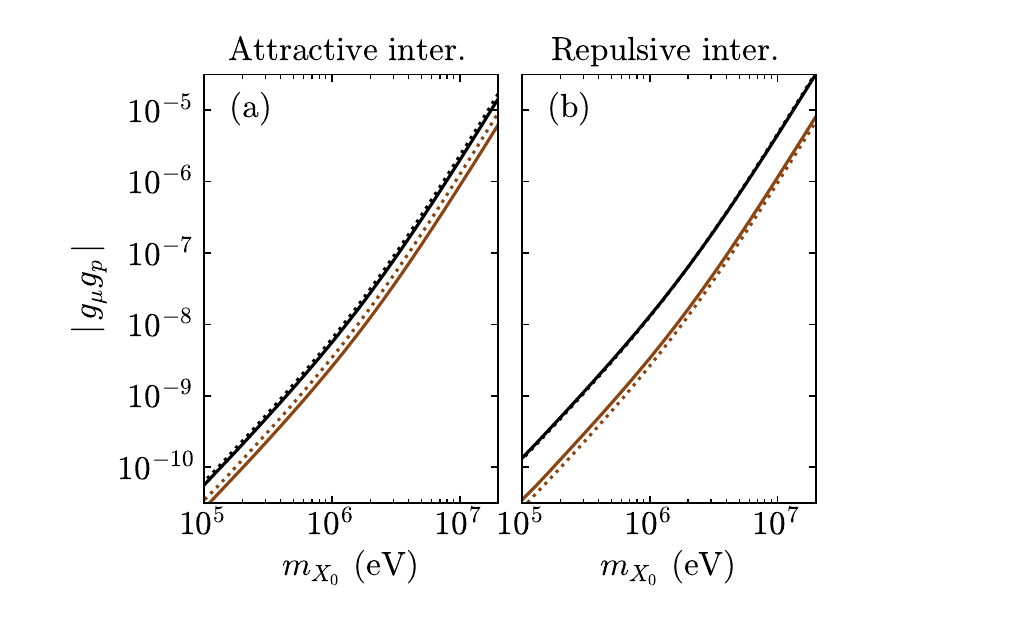}
\caption{The value of $|g_e g_p|$ above which an NP interaction is excluded at the 95\% confidence level, assuming that the ratio $g_d / g_p$ is either 2 (solid curves) or 0 (dotted curves) and that $g_\mu = g_e$, for a mediator mass above $1 \times 10^5$~eV. (a): Attractive NP interaction. (b): Repulsive NP interaction. Black curves: results based on the World spectroscopic data as in Fig.~\ref{fig:fig1_6boxes1p6}. Brown curves: results based only on the isotope shift measurements as in Fig.~\ref{fig:fig1_4boxes}.
}
\label{fig:highmasses}
\end{figure}
As is well known, however, the nuclear radii derived from the measurements in muonic hydrogen and muonic deuterium are in severe tension with much of the high precision data currently available for electronic hydrogen and electronic deuterium \cite{Bernauer2020}. These inconsistencies hamper calculations of bounds combining muonic and electronic species to the extent that our theoretical model cannot be made to match the whole set of data for any value of $g_eg_p$ when $g_d \not= g_p$. The model cannot be made to match the data for an attractive NP interaction, whereas a match with a relatively weak confidence level (up to 38\%) is found at non-zero values of $g_eg_p$ for a repulsive interaction (see Appendix~\ref{sec:furtherresults}). The better match found in the latter case illustrates the fact, already known \cite{Jones2020,Delaunay2023,Brandt2022}, that including a weak NP interaction in the theoretical model may improve its overall agreement with the World data. However, the relevance of this result is unclear given the residual inconsistencies within this data.

The CODATA group alleviated the difficulty of fitting the data to the Standard Model theory by increasing all the experimental uncertainties by 60\% \cite{Tiesinga2021}.
Doing so yields the bounds and distributions of confidence level presented in Fig.~\ref{fig:fig1_6boxes1p6}, for masses up to 100~keV, and in Fig.~\ref{fig:highmasses} (the black curves), for masses above 100~keV.
Comparing with Fig.~\ref{fig:fig1_10boxes1p0} shows that taking the muonic data into account tightens the bounds on NP further, by a considerable extent above 100~keV, despite the larger experimental uncertainties assumed in the calculation. (Results for an error magnification of 20\% rather than 60\% can be found in Appendix~\ref{sec:furtherresults}, for comparison.)

\begin{figure}[t!]
\centering
\includegraphics[width=\columnwidth]{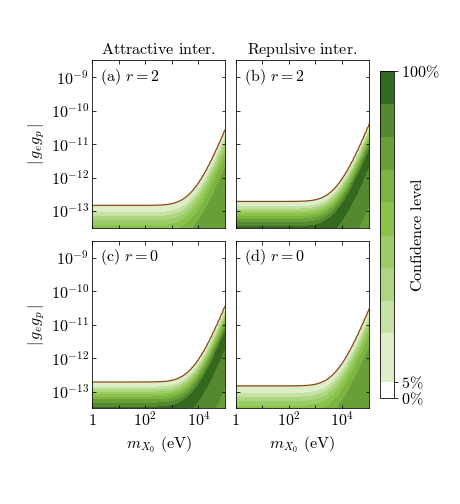}
$\mbox{}$\\[-1.0cm]
\caption{
Confidence level that an NP interaction is compatible with the 1s$_{1/2}$~--~2s$_{1/2}$ transition frequency measured in electronic hydrogen and electronic deuterium and with the nuclear charge radii derived from measurements in muonic hydrogen and muonic deuterium, assuming that the ratio $r = g_d/g_p$ is either 2 or 0 and that $g_\mu = g_e$. The experimental uncertainties are not changed here. The boxes on the left side of the figure refer to an attractive NP interaction, the others to a repulsive NP interaction. As in Fig.~\ref{fig:fig1_10boxes1p0}, the possibility of an NP interaction with parameters falling in a white region is excluded at the 95\% confidence level. The brown curves indicate the corresponding upper bounds on the value of $g_eg_p$.
}
\label{fig:fig1_4boxes}
\end{figure}
The difficulty of fitting the theoretical model to experiment can also be turned around without magnifying the experimental uncertainties by using only selected subsets of the available data. For $g_d \not= g_p$, the tightest bounds are obtained by combining the muonic values of the nuclear radii with the isotope shift of the 1s$_{1/2}$~--~2s$_{1/2}$ transition in the electronic species, owing to the particularly small experimental errors on these quantities \cite{noteaboutDelaunayetal}. 
The results for $r = 0$ or 2 are presented in Fig.~\ref{fig:highmasses} (the brown curves) and in Fig.~\ref{fig:fig1_4boxes}. They are also tabulated in Table~\ref{table:gmutoge} for $r =2$ (the column headed $g_\mu = g_e$). We do not present results for $r=1$, as this approach is unsuitable for this case (see Appendix~\ref{sec:othermethod}). 
Compared to the results of Figs.~\ref{fig:fig1_6boxes1p6}(c)-(f), the bounds obtained in this approach are significantly tighter. As in the case of the results based only on the electronic species, they vary with the ratio $r$ and strengthen rapidly when this ratio starts departing from 1 --- see the dashed brown curves in Fig.~\ref{fig:variable_ratio_2023}. We also note the sensitivity of these results to the details of small QED corrections. The theory of the Lamb shift in muonic deuterium has been further extended over the last few years \cite{Lensky2022,Pohl2016,Kalinowski2019}. As shown in Appendix~\ref{sec:furtherresults}, these new theoretical developments have significantly impacted on the conclusions one could draw from the isotope shift data in respect to the bounds on a hypothetical NP interaction.

\begin{table}
\caption{The value of $|g_eg_p|$ above which an NP interaction is excluded at the 95\% confidence level, for different values of
$m_{X_0}$ (the mass of the NP carrier) and different values of the ratio $g_\mu/g_e$, as calculated in the isotope shift approach for $g_d/g_p=2$. The numbers between square brackets indicate the powers of ten.}
\label{table:gmutoge}
\begin{center}
\begin{ruledtabular}
\begin{tabular}{lllll}
$m_{X_0}$ & $g_\mu = 0$ & $g_\mu = g_e$ & $g_\mu = 10\, g_e$ & $g_\mu = 100\, g_e$\\[1mm]
\tableline\\[-3mm]
\multicolumn{5}{c}{Attractive NP interaction} \\
$\leq 100$~eV & $1.5[-13]\;\;$ & $1.5[-13]\;\;$ & $1.5[-13]$ & $1.5[-13]$
\\
1~keV & $1.7[-13]$ & $1.7[-13]$ & $1.7[-13]$ &$1.7[-13]$
 \\
 10~keV & $6.8[-13]$ & $6.8[-13]$ & $6.8[-13]$ & $6.8[-13]$
 \\
 100~keV & $2.6[-11]$ & $2.6[-11]$ & $2.6[-11]$ & $2.7[-11]$
 \\
 1~MeV & $2.3[-9]\;\:$ & $2.6[-9]\;\:$ & $1.2[-7]\;\:$ & $3.3[-10]$
 \\
 10~MeV & $2.3[-7]\;\:$ & $8.8[-7]\;\:$ & $4.6[-8]\;\:$ & $4.1[-9]\;\:$
 \\[1mm]
 \multicolumn{5}{c}{Repulsive NP interaction} \\
$\leq 100$~eV & $1.9[-13]$ & $1.9[-13]$ & $1.9[-13]$ & $1.9[-13]$
\\
1~keV & $2.2[-13]$ & $2.2[-13]$ & $2.2[-13]$ &$2.2[-13]$
 \\
 10~keV & $8.9[-13]$ & $8.9[-13]$ & $8.9[-13]$ & $8.9[-13]$
 \\
 100~keV & $3.4[-11]$ & $3.4[-11]$ & $3.4[-11]$ & $3.5[-11]$
 \\
 1~MeV & $3.0[-9]\;\:$ & $3.4[-9]\;\:$ & $8.8[-8]\;\:$ & $2.5[-10]$
 \\
 10~MeV & $3.0[-7]\;\:$ & $1.2[-6]\;\:$ & $3.5[-8]\;\:$ & $3.1[-9]\;\:$
\end{tabular}
\end{ruledtabular}
\end{center}
\end{table}
How these bounds vary with the ratio $g_\mu/g_e$ is illustrated by Table~\ref{table:gmutoge}.
They do not significantly depend on the value of this ratio for mediator masses up to about 100~keV. At higher masses, however, they tend to move to higher values of $|g_eg_p|$ as $g_\mu/g_e$ increases, pass through a maximum and become tighter again for still higher values of this ratio. The maximum is reached at around or above 10~MeV for $g_\mu \approx g_e$, and at around 1~MeV for higher values of $g_\mu/g_e$. Except in that mass region,  taking $g_\mu = g_e$ in the calculation of these bounds is therefore a conservative assumption.

\section{Reach of the isotope shift approach}
\label{sec:isotopeshift}

The bounds on $g_eg_p$ discussed in the previous section are based on the current experimental evidence. We now consider the prospects for further tightening these bounds. Specifically, we look at how future isotope shift measurements could be used to this effect. We address this issue by using a method similar to that outlined in Appendix~\ref{sec:othermethod} for the specific case of the 1s$_{1/2}$~--~2s$_{1/2}$ interval, here generalized to other transitions. In effect, we also generalize the isotope shift calculations reported in Ref.~\cite{Delaunay2017}. 

\subsection{Bounds based on a single isotope shift}
\label{sec:single}

\subsubsection{Method}

The case of a single transition is particularly simple. Let us assume that a transition from a state $a$ to a state $b$ has been measured both in hydrogen and in deuterium, the result being the transition frequencies $\nu^{\rm exp}_{ba,{\rm eH}}$ and $\nu^{\rm exp}_{ba,{\rm eD}}$. We equate these two experimental transition frequencies to the corresponding theoretical transition frequencies, the latter including a possible NP shift:
\begin{align}
\nu^{\rm exp}_{ba,{\rm eH}} &=
{\cal R} \tilde{\nu}^{\rm g}_{ba,{\rm eH}} +
r_p^2\,\tilde{\nu}^{\rm ns}_{ba,{\rm eH}} +
\nu^{\rm oc}_{ba,{\rm eH}} + \nu^{\rm NP}_{ba,{\rm eH}}, \label{eq:H} \\
\nu^{\rm exp}_{ba,{\rm eD}}&=
{\cal R} \tilde{\nu}^{\rm g}_{ba,{\rm eD}} +
r_d^2\,\tilde{\nu}^{\rm ns}_{ba,{\rm eD}} +
\nu^{\rm oc}_{ba,{\rm eD}} + \nu^{\rm NP}_{ba,{\rm eD}}. \label{eq:D}
\end{align}
As above, ${\cal R}$ is the Rydberg frequency, $\nu^{\rm NP}_{{ba,\rm eH}}$ and $\nu^{\rm NP}_{ba,{\rm eD}}$  are NP shifts,
$r_p^2 \tilde{\nu}_{ba,{\rm eH}}^{\rm ns}$ and $r_d^2 \tilde{\nu}_{ba,{\rm eD}}^{\rm ns}$ are QED corrections proportional to the square of the respective nuclear radius, and
$\nu_{ba,{\rm eH}}^{\rm oc}$ and $\nu_{ba,{\rm eD}}^{\rm oc}$ are the sums of all the other SM corrections to the gross structure terms ${\cal R} \tilde{\nu}^{\rm g}_{ba,{\rm eH}}$ and
${\cal R} \tilde{\nu}^{\rm g}_{ba,{\rm eD}}$.
We will denote the corresponding isotope shifts by $\Delta\nu^{\rm exp}_{ba}$,
$\Delta\nu^{\rm g}_{ba}$, $\Delta\nu^{\rm ns}_{ba}$, $\Delta\nu^{\rm oc}_{ba}$ and
$\Delta\nu^{\rm NP}_{ba}$:
\begin{align}
 \Delta\nu^{\rm exp}_{ba} &=
    \nu^{\rm exp}_{ba,{\rm eD}} - \nu^{\rm exp}_{ba,{\rm eH}},\\
  \Delta\nu^{\rm g}_{ba} &= {\cal R}\tilde{\nu}^{\rm g}_{ba,{\rm eD}} - {\cal R}\tilde{\nu}^{\rm g}_{ba,{\rm eH}} \\
  \Delta\nu^{\rm ns}_{ba} &=
    r^2_{d}\, \tilde{\nu}^{\rm ns}_{ba,{\rm eD}} -
    r^2_{p}\, \tilde{\nu}^{\rm ns}_{ba,{\rm eH}}, \\
 \Delta\nu^{\rm oc}_{ba} &=
    \nu^{\rm oc}_{ba,{\rm eD}} - \nu^{\rm oc}_{ba,{\rm eH}},\\
  \Delta\nu^{\rm NP}_{ba} &=
    \nu^{\rm NP}_{ba,{\rm eD}} - \nu^{\rm NP}_{ba,{\rm eH}}.  
\end{align}
As given by Eq.~(\ref{eq:gross}), 
\begin{align}
\Delta\nu^{\rm g}_{ba} &= 
    {\cal R}\left[\frac{m_{\rm r}^{\rm eD}}{m_e} - \frac{m_{\rm r}^{\rm eD}}{m_e}\right]\left(\frac{1}{n_a^2} - \frac{1}{n_b^2}\right),
    \end{align}
where $m_{\rm r}^{\rm eH}$ ($m_{\rm r}^{\rm eD}$) is the reduced masses of electronic hydrogen (deuterium) and $m_e$ is the mass of the electron.  To conform with previous work on the isotope shift \cite{Jentschura2011,Pachucki2018}, and contrary to Section~\ref{sec:methods}, we only include the main nuclear size correction in $\Delta\nu^{\rm ns}_{ba}$. Thus
    \begin{align}
\Delta\nu^{\rm ns}_{ba} &= -
\frac{2\alpha^4 m_e c^2}{3h}\,
\left[
\left({m_{\rm r}^{\rm eD} \over m_e}\right)^3
{r_d^2 \over \lambdabar_{\rm C}^2}
-
\left({m_{\rm r}^{\rm eH} \over m_e}\right)^3
{r_p^2 \over \lambdabar_{\rm C}^2}
\right]\nonumber \\
& \qquad \qquad \qquad \qquad \qquad \times\left({\delta_{l_a,0} \over n_a^3} - {\delta_{l_b,0}\over n_b^3}\right),
\label{eq:Deltans}
\end{align}
and $\Delta\nu^{\rm oc}_{ba}$ includes all the other corrections to the gross structure term. 

We now use the fact that $\nu^{\rm NP}_{ba,{\rm eH}}$ is proportional to $g_eg_p$ and $\nu^{\rm NP}_{ba,{\rm eD}}$ to $g_eg_d$. Since $g_eg_d = r g_eg_p$ where $r$ is defined by Eq.~(\ref{eq:rdefined}), $\nu^{\rm NP}_{ba,{\rm eH}}$, $\nu^{\rm NP}_{ba,{\rm eD}}$ and ${\Delta}\nu^{\rm NP}_{ba}$ can be written in terms of $g_eg_p$-independent shifts $\tilde{\nu}^{\rm NP}_{ba,{\rm eH}}$,
$\tilde{\nu}^{\rm NP}_{ba,{\rm eD}}$ and ${\Delta}\tilde{\nu}^{\rm NP}_{ba}$. Namely,
\begin{align}
    \nu^{\rm NP}_{ba,{\rm eH}} &= 
    g_eg_p\,\tilde{\nu}^{\rm NP}_{ba,{\rm eH}} \\
    \nu^{\rm NP}_{ba,{\rm eD}} &= 
    r\, g_eg_p\,\tilde{\nu}^{\rm NP}_{ba,{\rm eD}},\\
    \Delta\nu^{\rm NP}_{ba} &= 
    g_eg_p\,\tilde{\nu}^{\rm NP}_{ba},
\end{align}
with
\begin{align}
    \Delta\tilde{\nu}^{\rm NP}_{ba} &= 
    r\tilde{\nu}^{\rm NP}_{ba,{\rm eD}} - \tilde{\nu}^{\rm NP}_{ba,{\rm eH}}.
\end{align}
Subtracting Eq.~(\ref{eq:H}) from Eq.~(\ref{eq:D}) and rearranging yields
\begin{equation}
    g_eg_p =
    \frac{\Delta\nu^{\rm exp}_{ba} - \Delta\nu^{\rm SM}_{ba}}{{\Delta}\tilde{\nu}^{\rm NP}_{ba}},
    \label{eq:identity}
\end{equation}
where
\begin{equation}
\Delta\nu^{\rm SM}_{ba} = 
    {\Delta}\nu^{\rm g}_{ba} +
    \Delta\nu^{\rm ns}_{ba}  + \Delta\nu^{\rm oc}_{ba}.
    \label{eq:deltath}
\end{equation}
Eq.~(\ref{eq:identity}) gives the strength of a hypothetical NP interaction which would explain a discrepancy between a measured transition frequency and the value predicted by the Standard Model. However, because of the measurements' finite precision and the  theoretical uncertainties in the QED corrections, this equation defines this strength only approximately. We denote the errors on $\Delta\nu^{\rm exp}_{ba}$ and
$\Delta\nu^{\rm SM}_{ba}$ by $\sigma^{\rm exp}_{ba}$ and $\sigma^{\rm SM}_{ba}$ and the total error on the numerator of Eq.~(\ref{eq:identity}) by $\sigma_{ba}$:
\begin{equation}
\sigma_{ba} = \sqrt{\left(\sigma^{\rm exp}_{ba}\right)^2 +  \left(\sigma^{\rm SM}_{ba}\right)^2}.
    \label{eq:errtot}
\end{equation}
The value of $g_eg_p$ predicted by Eq.~(\ref{eq:identity}) has a 95\% confidence interval of $\pm \sigma(|g_eg_p|)$, where \cite{shortnote}
\begin{equation}
    \sigma(|g_eg_p|) = 1.96 \,\sigma_{ba} / {\Delta}\tilde{\nu}^{\rm NP}_{ba}.
    \label{eq:deltagegp}
\end{equation}
It should be noted that $\sigma(|g_eg_p|)$ is not a bound on the strength of the NP interaction. Rather, $\sigma(|g_eg_p|)$ is half the difference between the most positive and negative values of $g_e g_p$ consistent with this interaction. The value of $\sigma(|g_eg_p|)$ thus indicates the sensitivity of the method. Eqs.~(\ref{eq:errtot}) and (\ref{eq:deltagegp}) show that this sensitivity is determined by the values of
$\sigma^{\rm exp}_{ba}$, $\sigma^{\rm SM}_{ba}$ and $\Delta\tilde{\nu}^{\rm NP}_{ba}$ for the interval considered. Since $\sigma(|g_eg_p|)$ is inversely proportional to ${\Delta}\tilde{\nu}^{\rm NP}_{ba}$ and ${\Delta}\tilde{\nu}^{\rm NP}_{ba}$ is particularly small for $g_d \approx g_p$, the method is unsuited to the $r=1$ case.

In regard to the measurements, we assume errors $\sigma^{\rm exp}_{ba,{\rm eD}}$ and $\sigma^{\rm exp}_{ba,{\rm eH}}$ on $\nu^{\rm exp}_{ba,{\rm eD}}$ and $\nu^{\rm exp}_{ba,{\rm eH}}$ taken individually, and an error $\sigma^{\rm exp}_{ba}$ on the difference $\nu_{ba,{\rm eD}}^{\rm exp} - \nu_{ba,{\rm eH}}^{\rm exp}$. Assuming that $\sigma^{\rm exp}_{ba,{\rm eD}}$ and $\sigma^{\rm exp}_{ba,{\rm eH}}$ are uncorrelated,
\begin{equation}
    \sigma^{\rm exp}_{ba} = \sqrt{\left(\sigma^{\rm exp}_{ba,{\rm eD}}\right)^2 +  \left(\sigma^{\rm exp}_{ba,{\rm eH}}\right)^2}.
    \label{eq:errexp}
\end{equation}
Regarding the Standard Model theory, we note that
\begin{equation}
    \sigma^{\rm SM}_{ba} = \sqrt{\left(\sigma^{\rm g}_{ba}\right)^2 +  \left(\sigma^{\rm ns}_{ba}\right)^2 +
    \left(\sigma^{\rm oc}_{ba}\right)^2},
    \label{eq:errSM}
\end{equation}
where $\sigma^{\rm g}_{ba}$, $\sigma^{\rm ns}_{ba}$ and $\sigma^{\rm oc}_{ba}$ are the errors on the gross structure term
$\Delta\nu^{\rm g}_{ba}$, on the nuclear size term $\Delta\nu^{\rm ns}_{ba}$, and on the other corrections term $\Delta\nu^{\rm oc}_{ba}$.

We first consider the error on the gross structure term. Using the CODATA~2018 values of the mass ratios $m_e/m_p$ and $m_e/m_d$ and the correlation in their uncertainties \cite{Tiesinga2021},
\begin{equation}
{m_{\rm r}^{\rm eD} \over m_e} - {m_{\rm r}^{\rm eH} \over m_e} =
2.71951069849(31) \times 10^{-4}.
\label{eq:massratiodiff}
\end{equation}
The relative error on this result, $1.1\times 10^{-10}$, is two orders of magnitude larger than the relative error on the Rydberg frequency \cite{Tiesinga2021}. The error on the gross structure term is therefore dominated by the error on the difference of mass ratios.
We find \cite{noteaboutmasses}
\begin{equation}
    \sigma^{\rm g}_{ba} = \mbox{0.10 kHz} \times 
    \left({1 \over n_a^2} - {1\over n_b^2}\right).
    \label{eq:sigmag}
\end{equation}

The error on the nuclear size term is dominated by the error on $r_p^2$ and, particularly, by the more significant error on $r_d^2$. The other quantities in Eq.~(\ref{eq:Deltans}) have a negligible error. We use the radii derived from the Lamb shift in muonic hydrogen and muonic deuterium, 
here re-calculated for taking into account a hypothetical NP interaction (we set $g_\mu = g_e$ when obtaining the numerical results presented in this section).
We found that the errors on these radii do not significantly vary with $g_eg_p$ in the range of values of $g_eg_p$ relevant for this work. They can be taken to be equal to the values determined without allowance for an NP interaction \cite{Antognini2013,Lensky2022}, which gives  
\begin{equation}
    \sigma^{\rm ns}_{ba} = \mbox{5.3 kHz} \times \left({\delta_{l_a,0} \over n_a^3} - {\delta_{l_b,0}\over n_b^3}\right).
\end{equation}

The error on the other corrections term can be derived from previous work \cite{Jentschura2011,Pachucki2018}. In particular, Pachucki, Patk\'{o}\v{s} and Yerokhin \cite{Pachucki2018} obtained 0.42~kHz for the total error on the theoretical 1s$_{1/2}$~--~2s$_{1/2}$ transition frequency, excluding errors that we include in $\sigma^{\rm ns}_{ba}$ or $\sigma^{\rm g}_{ba}$ in the present work  (i.e., the error on the leading nuclear size contribution and the error on the Dirac contribution to the isotope shift). Accordingly, we take $\sigma^{\rm oc}_{ba}$ to be 0.42~kHz for the 1s$_{1/2}$~--~2s$_{1/2}$ interval.
The main contribution to this error comes from the dominant QED corrections, which scale with the principal quantum number $n$ like $1/n^3$. Some subdominant corrections have a more complicated scaling with $n$; however, they are negligible in the present context. Hence, for transitions between two s-states, we set
\begin{equation}
    \sigma^{\rm oc}_{ba} = \mbox{0.48~kHz}\times
    \left({1 \over n_a^3} - {1\over n_b^3}\right),
    \label{eq:sigmaoc}
\end{equation}
which ensures that $\sigma^{\rm oc}_{ba} = 0.42$~kHz for $n_a = 1$ and $n_b =2$. For transitions between s-states, both $\sigma^{\rm g}_{ba}$ and $\sigma^{\rm oc}_{ba}$ are thus typically one order of magnitude smaller than $\sigma^{\rm ns}_{ba}$. Consequently, the method's sensitivity for such transitions is primarily limited by the experimental error and the uncertainty on the nuclear charge radii.

The error on the Lamb shift is considerably smaller for p- or d-states than for s-states of the same principal quantum number.
Setting $\sigma^{\rm oc}_{ba}$ equal to 0 for intervals not involving s-states is thus appropriate, which results in a smaller value of $\sigma_{ba}$ for the same experimental error. However, the gain in sensitivity arising from this smaller overall error may be partly or entirely negated by a smaller value of  $\Delta\tilde{\nu}^{\rm NP}_{ba}$ (transitions not involving s-states are necessarily transitions between two excited states, and the NP shift of such transitions is always smaller than the NP shift of a transition between the ground state and an excited state).

\subsubsection{Numerical illustration}

\begin{table}
\caption{The sensitivity parameter $\sigma(|g_eg_p|)$ for $m_{X_0}\lesssim 1$~eV, $g_d/g_p = 2$ and $g_\mu/g_e = 1$, as calculated from the isotope shift of each of the transitions indicated in the left column. The second column gives the experimental error assumed in the calculation. The theoretical limit specified in the last column is the value $\sigma(|g_eg_p|)$ for $\sigma_{ba}^{\rm exp}=0$. The numbers between square brackets indicate the powers of ten.}
\label{table:table2}
\begin{center}
\begin{ruledtabular}
\begin{tabular}{lccc}
Transition & $\sigma_{ba}^{\rm exp}$ & $\sigma(|g_eg_p|)$ & Th.~lim. \\[1mm]
\tableline\\[-3mm]
1s$_{1/2}$~--~2s$_{1/2}$ & 15~Hz & 1.7[-13]  & 1.7[-13] \\[1mm]
1s$_{1/2}$~--~3s$_{1/2}$ & 1~kHz &  1.6[-13] & 1.6[-13] \\[1mm]
1s$_{1/2}$~--~20s$_{1/2}$ & 1~kHz &  1.5[-13] & 1.5[-13] \\[1mm]
2s$_{1/2}$~--~20s$_{1/2}$ & 1~kHz & 1.4[-13] & 7.5[-14]\\
& 100~Hz & 7.6[-14] & 7.5[-14]\\
& 15~Hz & 7.5[-14] & 7.5[-14]\\[1mm]
8s$_{1/2}$~--~20s$_{1/2}$ & 100~Hz & 2.1[-13] & 2.1[-14]\\
& 15~Hz & 3.8[-14] & 2.1[-14]\\[1mm]
8d$_{5/2}$~--~20d$_{5/2}$ & 100~Hz & 2.1[-13]  & 2.8[-15] \\
& 15~Hz & 3.2[-14] & 2.8[-15]\\[1mm]
3d$_{5/2}$~--~20d$_{5/2}$ & 1~kHz & 2.6[-13] & 2.8[-15] \\
& 100~Hz & 2.6[-14] & 2.8[-15]\\[1mm]
\end{tabular}
\end{ruledtabular}
\end{center}
\end{table}
\begin{figure}[t!]
\centering
\includegraphics[width=0.95\columnwidth]{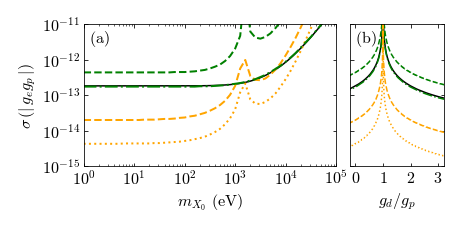}
\caption{The sensitivity parameter $\sigma(|g_eg_p|)$ for $g_\mu=g_e$, (a) vs.\ $m_{X_0}$, assuming that $g_d/g_p = 2$; (b) vs.\ the ratio $g_d/g_p$, assuming that $m_{X_0} = 1$~eV. Solid black curves: results based on the existing data for the isotope shift of the 1s$_{1/2}$~--~2s$_{1/2}$ interval. Dash-dotted green curves (almost indistinguishable from the solid black curves): results based on a hypothetical measurement of the isotope shift of the 1s$_{1/2}$~--~3s$_{1/2}$ interval with an experimental error of 1~kHz. Dashed green curves: the results which would be obtained by combining the isotope shift of the 1s$_{1/2}$~--~2s$_{1/2}$ interval with that of the 1s$_{1/2}$~--~3s$_{1/2}$ interval, assuming an experimental error of 1~kHz error on the latter. Dashed orange curves: the results which would be obtained by combining the isotope shift of the 1s$_{1/2}$~--~2s$_{1/2}$ interval with that of the 2s$_{1/2}$~--~20s$_{1/2}$ interval, assuming an experimental error of 100~Hz error on the latter. Dotted orange curves: the same as the dashed orange curves but for an experimental error of 15~Hz.
}
\label{fig:fig4}
\end{figure}

The potential reach of this approach in the low mass region is indicated by the values of $\sigma(|g_eg_p|)$ presented in Table~\ref{table:table2}. (Recall that this parameter determines the method's sensitivity to the effect of an NP interaction, with a lower value of $\sigma(|g_eg_p|)$ corresponding to a higher sensitivity.)  The values of $\sigma_{ba}^{\rm exp}$ assumed in the calculation are hypothetical, with the exception of the entry for the 1s$_{1/2}$~--~2s$_{1/2}$ interval, which is the experimental uncertainty already achieved for this transition \cite{Parthey2010}. The right-hand most column of the table gives the value of $\sigma(|g_eg_p|)$ calculated for $\sigma_{ba}^{\rm exp}=0$.
Values of $\sigma(|g_eg_p|)$ based on actual or hypothetical measurements of the isotope shift of the 1s$_{1/2}$~--~2s$_{1/2}$ and 1s$_{1/2}$~--~3s$_{1/2}$ transition frequencies are also presented in Fig.~\ref{fig:fig4} over a wide range of mediator masses.

The first row of Table~\ref{table:table2} and the solid black curves in the figure refer to the value of $\sigma(|g_eg_p|)$ based on the existing measurement of the isotope shift of the 1s$_{1/2}$~--~2s$_{1/2}$ interval \cite{Parthey2010}. This value is entirely consistent with the results of Fig.~\ref{fig:fig1_4boxes} and with the bounds calculated in Appendix~\ref{sec:othermethod}, which are based on the same data. 
The second row in the table and the dash-dotted green curves in Fig.~\ref{fig:fig4} (almost indistinguishable from the solid black curves) refer to the value of $\sigma(|g_eg_p|)$ calculated from a hypothetical measurement of the isotope shift of the 1s$_{1/2}$~--~3s$_{1/2}$ interval, assuming an experimental uncertainty of $1$~kHz on this quantity (for comparison, the error on the most recent determination of the 1s$_{1/2}$~--~3s$_{1/2}$ transition frequency in hydrogen is 0.72~kHz \cite{Grinin2020}). Although $\sigma_{ba}^{\rm exp}$ is considerably larger here, the resulting values of $\sigma(|g_eg_p|)$ are practically the same as those based on the 1s$_{1/2}$~--~2s$_{1/2}$ interval. The similarity between these two sets of results arises from the fact that $\sigma_{ba}$ --- the numerator of Eq.~(\ref{eq:deltagegp}) --- is dominated by the nuclear size error for these transitions, which is roughly the same for the two intervals: in relative terms, the larger value of $\sigma_{ba}^{\rm exp}$ only produces a small increase in the value of $\sigma_{ba}$, which is largely compensated by an increase in the value of $\Delta\tilde{\nu}^{\rm NP}_{ba}$ ($\Delta\tilde{\nu}^{\rm NP}_{ba}$ is almost 20\% larger for the 1s$_{1/2}$~--~3s$_{1/2}$ interval). Reducing the experimental uncertainties further would not improve the method's sensitivity for these two transitions. We note, however, that complementing the existing measurement of the 1s$_{1/2}$~--~3s$_{1/2}$ transition frequency in hydrogen by a measurement of its isotope shift would be useful as an independent check of the results derived from the 1s$_{1/2}$~--~2s$_{1/2}$ interval.

The scope for achieving a lower value of $\sigma(|g_eg_p|)$ within this approach can be inferred from the remaining rows of Table~\ref{table:table2}. Using transitions between the ground state and more highly excited states would not lead to a significant gain in sensitivity without significantly reducing the uncertainties on the nuclear charge radii. Using transitions from the metastable 2s$_{1/2}$ state to Rydberg states could lead to a twofold increase in sensitivity as long as the experimental error would not be much larger than 100~Hz (we have previously noted that reducing the experimental error to this level is likely to be achievable for such transitions \cite{Jones2020}). Reaching higher sensitivities would require ultra-high precision isotope shift measurements on more highly excited states. As is explained in Appendix~\ref{sec:limit}, the theoretical limit of $\sigma(|g_eg_p|)$ is as small as $2.8 \times 10^{-15}$ for transitions not involving s-states, but approaching this limit would be particularly challenging experimentally. It is worth noting that the 3d$_{3/2}$ and 3d$_{5/2}$ states have a more significant NP shift than the other d-states, which makes transitions from a 3d state to a Rydberg d-state of particular interest in this context. As seen from the table, a six-fold reduction of $\sigma(|g_eg_p|)$ from its current best value could be achieved if the isotope shift of, e.g., the 3d$_{5/2}$~--~20d$_{5/2}$ interval were measured with a precision of 100~Hz.

The other curves plotted in Fig.~\ref{fig:fig4} refer to results obtained by combining two different isotope shifts in an approach described in the next section.

\subsection{Bounds based on two isotope shifts}
\label{sec:pair}

As noted in Section~\ref{sec:single}, the sensitivity of the single-transition method is limited by the error on the nuclear size term when applied to transitions between s-states. However, as we now discuss, this error can be eliminated by combining the isotope shifts of two different intervals.

Let us imagine that experimental values of the isotope shift would be known for two different transitions, say, for a transition from a state $a$ to a state $b$ and for a transition from a state $c$ to a state $d$, all s-states. Eqs.~(\ref{eq:identity}) and (\ref{eq:deltath}) give
\begin{align}
    g_eg_p\,\Delta\tilde{\nu}^{\rm NP}_{ba} &=
    \Delta\nu^{\rm exp}_{ba} - 
    {\Delta}\nu^{\rm g}_{ba} -
    \Delta\nu^{\rm ns}_{ba}  - \Delta\nu^{\rm oc}_{ba}, \label{eq:first} \\
    g_eg_p\,\Delta\tilde{\nu}^{\rm NP}_{dc} &=
    \Delta\nu^{\rm exp}_{dc} - 
    {\Delta}\nu^{\rm g}_{dc} -
    \Delta\nu^{\rm ns}_{dc}  - \Delta\nu^{\rm oc}_{dc}. \label{eq:second}
\end{align}
(Recall that these equations refer to isotope shifts, not to transitions for a single isotope.) Let
\begin{align}
    \epsilon_1 = \frac{1}{n_a^3} - \frac{1}{n_b^3}, \qquad
    \epsilon_2 = \frac{1}{n_c^3} - \frac{1}{n_d^3}.
\end{align}
In view of Eq.~(\ref{eq:Deltans}), multiplying Eq.~(\ref{eq:second}) by $\epsilon_1/\epsilon_2$ and subtracting the product from Eq.~(\ref{eq:first}) cancels the nuclear size terms. The result can be written as follows:
\begin{align}
    g_eg_p &=
    \frac{\Delta^{\rm exp}_{ba,dc} - 
    {\Delta}^{\rm g}_{ba,dc} -
    {\Delta}^{\rm ns}_{ba,dc} -
    \Delta^{\rm oc}_{ba,dc}}{\tilde{\Delta}^{\rm NP}_{ba,dc}},
    \label{eq:identity2}
\end{align}
with
\begin{align}
    {\Delta}^{\rm exp}_{ba,dc} &= {\Delta}^{\rm exp}_{ba} - (\epsilon_1/\epsilon_2){\Delta}^{\rm exp}_{cd}, \\
    {\Delta}^{\rm g}_{ba,dc} &= {\Delta}^{\rm g}_{ba} - (\epsilon_1/\epsilon_2){\Delta}^{\rm g}_{cd}, \\
    {\Delta}^{\rm ns}_{ba,dc} &= {\Delta}^{\rm ns}_{ba} - (\epsilon_1/\epsilon_2){\Delta}^{\rm ns}_{cd}, \\
    {\Delta}^{\rm oc}_{ba,dc} &= {\Delta}^{\rm oc}_{ba} - (\epsilon_1/\epsilon_2){\Delta}^{\rm oc}_{cd}, \\
\end{align}
and
\begin{equation}
    \tilde{\Delta}^{\rm NP}_{ba,dc} = \tilde{\Delta}^{\rm NP}_{ba} - (\epsilon_1/\epsilon_2)\tilde{\Delta}^{\rm NP}_{cd}.
\end{equation}
We denote the uncertainties on
${\Delta}^{\rm exp}_{ba,dc}$, ${\Delta}^{\rm g}_{ba,dc}$, ${\Delta}^{\rm ns}_{ba,dc}$ and ${\Delta}^{\rm oc}_{ba,dc}$ by ${\sigma}^{\rm exp}_{ba,dc}$, ${\sigma}^{\rm g}_{ba,dc}$, ${\sigma}^{\rm ns}_{ba,dc}$ and ${\sigma}^{\rm oc}_{ba,dc}$, respectively.
Assuming that the individual experimental errors $\sigma^{\rm exp}_{ba}$ and $\sigma^{\rm exp}_{dc}$ are uncorrelated,
\begin{equation}
    \sigma^{\rm exp}_{ba,dc} = \sqrt{\left(\sigma^{\rm exp}_{ba}\right)^2 +
    \left(\epsilon_1\sigma^{\rm exp}_{dc}/\epsilon_2\right)^2}. 
\end{equation}
Given Eq.~(\ref{eq:sigmag}) and the fact that the error on the mass ratios contributes exactly in the same way for the two transitions,
\begin{equation}
    \sigma^{\rm g}_{ba,dc} = \mbox{0.10 kHz} \times \left[ \frac{1}{n_a^2} - \frac{1}{n_b^2} -
    \frac{\epsilon_1}{\epsilon_2}\left(\frac{1}{n_d^2} - \frac{1}{n_c^2} \right)\right].
\end{equation}
By construction, ${\Delta}^{\rm ns}_{ba,dc} =0$.
The errors on $\Delta^{\rm ns}_{ba}$ and $\Delta^{\rm ns}_{dc}$ being perfectly correlated, $\sigma^{\rm ns}_{ba,dc} $ is also zero. 
Given how the different contributions to the Lamb shift scale with $n$, and in the current absence of a more in depth study of this issue, we tentatively set 
\begin{equation}
    \sigma^{\rm oc}_{ba,dc} = \mbox{50~Hz},
\end{equation}
which is conservative.
The 95\% confidence interval of the value of $g_eg_p$ predicted by Eq.~(\ref{eq:identity2}) is thus $\pm \sigma(|g_eg_p|)$ with
\begin{equation}
    \sigma(|g_eg_p|) = 1.96 \, \,\frac{\sqrt{\left(\sigma^{\rm exp}_{ba,dc}\right)^2 + \left(\sigma^{\rm g}_{ba,dc}\right)^2+ \left(\sigma^{\rm oc}_{ba,dc}\right)^2 }}
    {|\tilde{\Delta}^{\rm NP}_{ba,dc}|}.
    \label{eq:deltagegp2}
\end{equation}

Compared to Eq.~(\ref{eq:deltagegp}), the numerator of Eq.~(\ref{eq:deltagegp2}) will typically be significantly smaller for transitions between s-states. However, the NP shifts of the two intervals also partly cancel, and therefore the denominator of Eq.~(\ref{eq:deltagegp2}) may also be significantly smaller, offsetting the gain made by reducing the magnitude of the numerator. 
The result may be a reduction of sensitivity (a larger value of $\sigma(|g_eg_p|)$) compared to the single transition method. For example, the value of $\sigma(|g_eg_p|)$ obtained by combining
the existing results for the 1s$_{1/2}$~--~2s$_{1/2}$ interval with the hypothetical results for the 1s$_{1/2}$~--~3s$_{1/2}$ interval is higher than the values obtained from each of these intervals taken individually (the result of the combined calculation is represented by a dashed green curve in Fig.~\ref{fig:fig4}).

Yet, a significant improvement in sensitivity on the single isotope shift method could be obtained by combining appropriately chosen intervals, provided the experimental uncertainties would be small enough. For example, the value of $\sigma(|g_eg_p|)$ based on the existing results for the 1s$_{1/2}$~--~2s$_{1/2}$ isotope shift would be reduced ten-fold by combining these results with measurements of the 2s$_{1/2}$~--~20s$_{1/2}$ isotope shift if the latter was obtained with an experimental uncertainty of about 100~Hz --- see the dashed orange curve in Fig.~\ref{fig:fig4}. Reducing this experimental uncertainty to 15~Hz would lower $\sigma(|g_eg_p|)$ to about $4.3 \times 10^{-15}$ in the low mass region (the dotted orange curve in the figure), which is close to the limit set by the current uncertainties on the relevant QED corrections and nuclear masses ($3.1 \times 10^{-15}$ for this transition).

Finally, we should emphasize that the results presented in Fig.~\ref{fig:fig4} depend on the value assumed for the ratio $g_d/g_p$, like those presented in Section~\ref{sec:results}. The values of $\sigma(|g_eg_p|)$ given by Eqs.~(\ref{eq:deltagegp}) and (\ref{eq:deltagegp2}) are roughly proportional to $1/|g_d/g_p -1|$. Setting $g_d = 2 g_p$ leads to the results shown in part (a) of Fig.~\ref{fig:fig4}. However, much larger values of  $\sigma(|g_eg_p|)$ would be found for $g_d \approx g_p$, as shown by part (b) of the figure.

\section{Conclusions}
\label{sec:conclusions}

Precision transition measurements in deuterium and hydrogen are a direct way to tension well-motivated extensions of the Standard Model. The results presented in this article illustrate the advantages of combining results for these two isotopes when setting bounds on a New Physics interaction that couples differently to a deuteron than to a proton.

Specifically, we have presented bounds based on the World spectroscopic hydrogen and deuterium data and bounds based only on the isotope shift of the 1s$_{1/2}$~--~2s$_{1/2}$ interval.
The former are more blunt than the latter for models in which $g_d \not= g_p$ (as defined above, $g_d$ and $g_p$ denote the coupling constants of the New Physics interaction with, respectively, a deuteron and a proton). However, they have the advantage of being based on a number of independent measurements and a broad range of transitions. Their scope is limited by their well known internal inconsistencies, though. Resolving those would be of benefit in strengthening these bounds, besides reducing the uncertainties on the values of fundamental constants \cite{Tiesinga2021}.
Where comparison is possible, our results are in broad agreement with those of Ref.~\cite{Delaunay2023}, which were obtained independently and focus on specific theoretical models. 

The bounds based on the isotope shift of the 1s$_{1/2}$~--~2s$_{1/2}$ interval are more stringent \cite{Delaunay2017}.
As explained in Section~\ref{sec:isotopeshift},
it would be useful to complement the existing measurements of the 1s$_{1/2}$~--~3s$_{1/2}$ interval in hydrogen by measurements in deuterium at a similar level of precision, so as to obtain bounds based on the isotope shift of this interval. Those bounds would provide a useful independent check on the results derived from the 1s$_{1/2}$~--~2s$_{1/2}$ isotope shift. Measurements of that transition in deuterium at the required level of precision are currently considered \cite{Yzombard2023}.

The sensitivity of the isotope shift method for these two intervals is limited by the current uncertainties on the theory and the nuclear charge radii. We found that a measurement of the isotope shift of the 2s$_{1/2}$~--~20s$_{1/2}$ interval (or more generally 2s$_{1/2}$ to Rydberg s-state) would make it possible to bypass this limitation, provided the experimental error would be sufficiently small --- e.g., of the order of  0.1~kHz or better for the 2s$_{1/2}$~--~20s$_{1/2}$ interval. As described above, theoretical errors can indeed be reduced by combining the isotope shift of such intervals and that of the 1s$_{1/2}$~--~2s$_{1/2}$ interval, making the experimental error the main limitation of the method. Compared to the results based on currently available spectroscopic measurements, achieving an experimental uncertainty of 0.1~kHz would improve the sensitivity of the method by one order of magnitude in the low mass region, for $g_d \not= g_p$.

These results and those of Ref.~\cite{Delaunay2023} thus suggest that any future precision experiments in hydrogenic atoms should consider whether the experimental method may be extended to deuterium as part of the planning stage.

\acknowledgements

We gratefully acknowledge helpful communications received from C\'{e}dric Delaunay and Yotam Soreq about their work in the area of this paper, and from Krzysztof Pachucki about the 2018 adjustment of the fundamental constants by the CODATA group.

\appendix

\section{Impact of an NP~interaction on the determination of $r_p$ and $r_d$ in muonic species}
\label{sec:rprd}

The nuclear charge radii derived from the experiments on muonic hydrogen and muonic deuterium were calculated from the respective Lamb shifts, which were themselves calculated from measured energy differences between hyperfine components of the 2s$_{1/2}$ and 2p$_{3/2}$ states \cite{Antognini2013,Pohl2016}. In the following, we ascertain the maximal values of the coupling constants $g_\mu g_p$ and $g_\mu g_d$ for which a hypothetical NP shift of the 2s$_{1/2}$~--~2p$_{3/2}$ interval  would have a negligible impact on these charge radii. We use atomic units throughout this appendix, except where specified otherwise.

The NP shift in question reduces to $\delta E_{21}^{\rm NP} - \delta E_{20}^{\rm NP}$ in the non-relativistic approximation, with $\delta E_{21}^{\rm NP}$ and $\delta E_{20}^{\rm NP}$ defined by Eq.~(\ref{eq:shift1}). From Appendix~B of paper~I, 
\begin{align}
 \delta E_{21}^{\rm NP} - \delta E_{20}^{\rm NP} 
    &= \frac{B}{2m_{\rm r}} \, \frac{C^2} {(C/m_{\rm r} + 1)^4},
    \label{eq:nonrelativistic}
\end{align}
where $m_{\rm r}$ is the reduced mass of the system ($m_{\rm r} \approx 186\,m_e$ for muonic hydrogen and $m_{\rm r} \approx 196\,m_e$ for muonic deuterium). When $C \ll 1$, this shift is thus smaller by a factor $m_e/m_{\rm r}$ for the muonic species as compared to the electronic species. According to Eq.~(\ref{eq:nonrelativistic}), it also goes rapidly to zero 
for $C \rightarrow 0$, i.e., in the low mass limit. However, this is so only because Eq.~(\ref{eq:nonrelativistic}) neglects relativistic effects and the spatial extension of the nucleus \cite{noteaboutrelativisticwf}.

Relativistic effects can be taken into account by replacing the radial wave functions $R_{nl}(r)$ by solutions of the Breit equation or the Dirac equation. We use Dirac wave functions and replace Eq.~(\ref{eq:shift1}) by
\begin{align}
   &\delta E^{{\rm NP}}_{n\kappa} 
    =  \nonumber \\ &\qquad
    \int_0^\infty \left[\big|P^{}_{n\kappa}(r)\big|^2 + \big|Q^{}_{n\kappa}(r)\big|^2\right]
    \frac{B\exp(-Cr)}{r}\,
    {\rm d}r, \label{eq:reldelta}
    \end{align}
where 
$P^{}_{n\kappa}(r)$ and $Q^{}_{n\kappa}(r)$ are the radial parts of the large and small components of the respective 4-component spinor multiplied by $r$,  and $\kappa$ is the relativistic quantum number ($\kappa = -1$ for the 2s$_{1/2}$ state, $1$ for the 2p$_{1/2}$ state, and $-2$ for the 2p$_{3/2}$ state). The functions $P^{}_{n\kappa}(r)$ and $Q^{}_{n\kappa}(r)$ are calculated for a particle of mass $\mu$ moving in a central potential well $V(r)$ and are
normalized according to the usual prescription,
\begin{equation}
    \int_0^\infty \left[\big|P^{}_{n\kappa}(r)\big|^2 + \big|Q^{}_{n\kappa}(r)\big|^2\right]
    \, {\rm d}r = 1.
\end{equation}

Setting $V(r) \equiv -1/r$ yields the Dirac wave functions of the respective hydrogenic states.
However, choosing $V(r)$ in this way would involve neglecting vacuum polarization, which in muonic hydrogen and muonic deuterium accounts for a large part of the Lamb shift \cite{Borie1982,Antognini2013b,Krauth2016}. It would also involve neglecting the finite extension of its charge distribution, which can be expected to play a larger role here than in normal hydrogen since the wave functions of muonic species are more compact. 
It is therefore preferable to set
\begin{equation}
    V(r) = V_{\rm Uehl}(r) + V_{N}(r)
\end{equation}
when estimating the NP shift. $V_{\rm Uehl}(r)$ is the Uehling potential, which accounts for the bulk of the vacuum polarization effects in muonic species \cite{Borie1982}, and $V_{N}(r)$ is the electrostatic muon-nucleus potential obtained in a model of the charge distribution of the latter. For convenience, we use a Gaussian model for this charge distribution, with an rms radius $r_N$ equal to the experimental charge radius of either the proton or the deuteron. Thus
\begin{equation}
    V_N(r) = -\frac{1}{r}\,
    \mbox{erf}\,(r/r_0),
\end{equation}
where $\mbox{erf}(\cdot)$ denotes the error function and $r_0 = (2/3)^{1/2}r_{N}$ \cite{Friar1979,Indelicato2013}. The Uehling potential is calculated for this Gaussian charge distribution rather than for a point nuclear charge, and both $V_N(r)$ and $V_{\rm Uehl}(r)$ are taken into account to all orders by solving the Dirac equation non-perturbatively. The approach follows Ref.~\cite{Indelicato2013}. We used the program {\tt QEDMOD} \cite{Shabaev2018} for calculating the Uehling potential and the program {\tt RADIAL} \cite{Salvat2019} for integrating the Dirac equation.

As mentioned at the beginning of this appendix, we want to ascertain how strong an NP interaction could be without significantly affecting the nuclear charge radii derived from the muonic data. We define an upper bound on $|g_\mu g_N|$, $|g_\mu g_N|_{\rm max}$, such that the effect of an NP interaction would be negligible if $|g_\mu g_N| < |g_\mu g_N|_{\rm max}$ but might be significant if $|g_\mu g_N| \geq |g_\mu g_N|_{\rm max}$. Conservatively, we take $|g_\mu g_N|_{\rm max}$ to be the value of $|g_\mu g_N|$ at which the NP shift of the 2s$_{1/2}$~--~2p$_{3/2}$ interval is 5\% of the error on the experimental Lamb shift --- i.e., 0.12~$\mu$eV in muonic hydrogen ($\mu$H) and 0.17~$\mu$eV in muonic deuterium ($\mu$D).
 
\begin{figure}[b!]
\centering
\includegraphics[width=0.95\columnwidth]{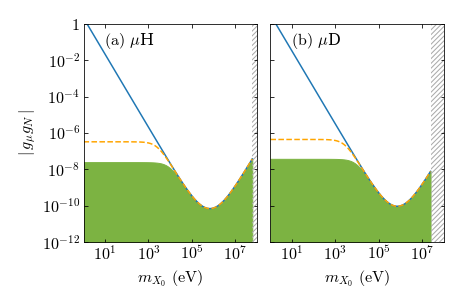}
\caption{Green areas: regions of the $(m_{X_0},|g_\mu g_N|)$ plane in which an NP interaction between the muon and the nucleus would not shift the measured 2s$_{1/2}$~--~2p$_{3/2}$ interval by more than 5\% of the experimental error on the respective Lamb shift in (a) muonic hydrogen ($g_N = g_p$); (b) muonic deuterium ($g_N = g_d$). These regions would be predicted to extend upwards to the dashed orange curves if vacuum polarization and the finite size of the nucleus were ignored, or to the solid
blue curves in the non-relativistic approximation for a point nucleus. We do not present results for the ranges of values of $m_{X_0}$ indicated by the shaded regions in view of the uncertainty on the form of the NP interaction within the nucleus.}
\label{fig:fig2}
\end{figure}
The region of the $(m_{X_0}, |g_\mu g_N|)$-plane in which a hypothetical NP interaction would be negligible according to this definition
is identified by the green areas in Fig.~\ref{fig:fig2}. We assume no definite relationship between $g_\mu g_N$ for $\mu$H and $g_\mu g_N$ for $\mu$D.
We stress that these green areas only indicate the region of the parameter space in which an NP shift would not matter for the deduction of the proton or deuteron radius; they do not indicate whether a hypothetical NP interaction might or might not be compatible with existing data.
Given the uncertainty on the form of the NP interaction inside the nucleus, we do not venture predictions of the importance of an NP shift for a mediator mass large enough that the region $r \leq 2 r_N$ contributes significantly to the integral in Eq.~(\ref{eq:reldelta}). This mass region starts at about 70~MeV for muonic hydrogen and about 30~MeV for muonic deuterium. The corresponding regions of Fig.~\ref{fig:fig2} are shaded to indicate that they are not considered in the present analysis.

Fig.~\ref{fig:fig2}(a) shows that $|g_\mu g_N|_{\rm max}$ does not much exceed $1\times 10^{-8}$ for $m_{X_0} < 10$~keV.
The data for electronic hydrogen exclude the possibility that $|g_e g_p|$ could be as large as $1\times 10^{-8}$ between 1~eV and 10~keV --- see, e.g.\ Figs.~\ref{fig:fig1_10boxes1p0}(a) and (b). Assuming that lepton universality still holds true for the coupling with an NP interaction, so that $g_\mu = g_e$, we can thus conclude that an NP interaction would have but a negligible impact on the values of $r_p$ in this mass range. Given the results of Figs.~\ref{fig:fig1_10boxes1p0}(c) and (d), the same also applies to the determination of $r_d$ from the measurements in muonic deuterium, at least for carrier masses between 10~eV and 10~keV. However, the bounds based on the spectroscopy of electronic hydrogen and electronic deuterium do not clearly exclude the possibility of a significant NP interaction with carrier mass well above 10~keV.

It is worth noting the importance of $V_{\rm Uehl}(r)$ and $V_N(r)$ in this context. Indeed, setting $V(r) = -1/r$ yields the result indicated by dashed orange curves in Figs.~\ref{fig:fig2}(a) and (b). As seen from these figures, $|g_\mu g_N|_{\rm max}$ is somewhat overestimated in the low mass region, within this approximation.

The value of $|g_\mu g_N|_{\rm max}$ derived from Eq.~(\ref{eq:nonrelativistic}) is also shown in Fig.~\ref{fig:fig2} (the solid blue curves): as is readily seen, the non-relativistic approximation is good for $m_{X_0} > 10$~keV and unsuitable for $m_{X_0} < 10$~keV. 

\mbox{}

\section{Notes about the calculations}
\label{sec:notes}

The results presented in this paper are largely based on the set of 29 measured transition energies between states of electronic hydrogen and electronic deuterium used by CODATA in their most recent determination of the Rydberg constant \cite{Tiesinga2021}. This set is listed in Table~X of that reference, as the input data A1 to A29. The set of data used in the present work differs in the following ways:
\begin{enumerate}
\item We also use the recent measurements of the 1s$_{1/2}$~--~3s$_{1/2}$ and 2s$_{1/2}$~--~8d$_{5/2}$ transition energies in electronic hydrogen \cite{Grinin2020,Brandt2022}, as well as the experimental results quoted in Ref.~\cite{DeVries2002} for transitions between high circular states of electronic hydrogen (specifically, and as explained in paper~I, we derive an experimental energy for the transition between the $(n=27,l=26)$ and $(n=28,l=27)$ states from the value of ${\cal R}$ obtained in that work).
\item For the input datum A28 (the 2s$_{1/2}$~--~2p$_{1/2}$ Lamb shift measurement of Ref.~\cite{Lundeen1981}), we use the value recently recommended in Ref.~\cite{Marsman2018} rather than the original value.
\item We exclude the input datum A7 (the 2013 measurement of the 1s$_{1/2}$~--~2s$_{1/2}$ transition energy in electronic hydrogen
\cite{Matveev2013}), in view of its strong correlation with the input datum A6 (the 2011 measurement of that transition).
This strong correlation tends to inflate the value of $\chi^2$ significantly in fits using both values.
When used individually, the resulting confidence levels are practically indistinguishable. However, they tend to decrease significantly when both these transition energies are included in the fit.
The difficulty does not arise for the other intervals for which several different experimental values are included in the data set, as in these other cases the experimental errors are not strongly correlated with each other.
\item We use the input datum A5 (the isotope shift of the 1s$_{1/2}$~--~2s$_{1/2}$ interval in electronic hydrogen \cite{Parthey2010}) as an independent input to the $\chi^2$ fit only when combining hydrogen and deuterium data. We do not use this datum at all in calculations of bounds based only in measurements in hydrogen. For bounds based only in measurements in deuterium, we use it in conjunction with the input datum A6 to generate an experimental value for the 1s$_{1/2}$~--~2s$_{1/2}$ transition frequency in deuterium \cite{Pohl2017}, which is used in the fit.  
\end{enumerate}
The experimental data for the muonic species are the Lamb shifts measured by the CREMA collaboration \cite{Antognini2013,Pohl2016}. 
The correlation coefficients used in the $\chi^2$-fitting are taken from Ref.~\cite{Tiesinga2021} (the errors on the experimental results not considered in that reference can safely be assumed to be uncorrelated with those considered in it and with each other). No magnification of the experimental uncertainties was made for producing the results shown in Figs.~\ref{fig:fig1_10boxes1p0}, \ref{fig:variable_ratio_2023} and \ref{fig:fig1_4boxes} and the results represented by the brown curves in Fig.~\ref{fig:highmasses}.
The experimental uncertainties were increased by 60\% for producing the results shown in Fig.~\ref{fig:fig1_6boxes1p6} and those represented by the black curves in Fig.~\ref{fig:highmasses}.

The Standard Model calculations follow Appendix~C of Paper~I for the electronic species and, except where specified otherwise, Refs.~\cite{Antognini2013} and \cite{Lensky2022} for the muonic species. We stress that we take the finite size of the nucleus into account both for the electronic and the muonic species, which is important for the accuracy of the models.

\section{Isotope shift analysis with an NP interaction}
\label{sec:othermethod}

As discussed in Section~\ref{sec:allres} and in previous work \cite{Delaunay2017}, tight bounds on the NP coupling constants can be derived from the measurements of the 1s$_{1/2}$~--~2s$_{1/2}$ interval in electronic hydrogen and deuterium, in conjunction with the measurements on the nuclear radius in muonic hydrogen and deuterium. The calculations reported in Section~\ref{sec:allres} follow a general approach applicable to any number of transitions. However, they do not make use of particularly precise theoretical results, reported in Refs.~\cite{Jentschura2011,Pachucki2018}, which are specific to the isotope shift of that interval. An alternative approach to obtaining bounds on an NP interaction, taking advantage of these theoretical results, is outlined in this appendix.

The key experimental evidence is the difference,  $\Delta \nu^{\rm exp}$, between the 1s$_{1/2}$~--~2s$_{1/2}$ transition frequency of electronic deuterium and that of electronic hydrogen \cite{Parthey2010}:
\begin{align}
\Delta \nu^{\rm exp} &= \Delta \nu_{{\rm 2s1s},e{\rm D}}^{\rm exp} - \Delta \nu_{{\rm 2s1s},e{\rm H}}^{\rm exp} \nonumber \\
&= 670\,994\,334.606(15)~\mbox{kHz}.
\end{align}
Allowing for a hypothetical NP interaction, the theoretical isotope shift is
$\Delta \nu^{\rm SM} + \Delta \nu^{\rm NP}$, where $\Delta \nu^{\rm SM}$ is the isotope shift calculated within the Standard Model and $\Delta \nu^{\rm SM}$ is a New Physics correction. In terms of the NP shifts defined by Eq.~(\ref{eq:DeltaNPdefined}),
\begin{align}
\Delta \nu^{\rm NP} &= \Delta \nu_{{\rm 2s1s},e{\rm D}}^{\rm NP} - \Delta \nu_{{\rm 2s1s},e{\rm H}}^{\rm NP}.
\end{align}
Moreover \cite{Pachucki2018},
\begin{align}
\Delta \nu^{\rm SM} &=
 \Delta \nu_{{\rm 2s1s},e{\rm D}}^{\rm SM} - \Delta \nu_{{\rm 2s1s},e{\rm H}}^{\rm SM} \nonumber \\
&=
670\,999\,567.88(42)~\mbox{kHz}  \nonumber \\
&\qquad - \frac{7\alpha^4 m_e c^2}{12 h\lambdabar_{\rm C}^2} 
\left[\left(\frac{m_{\rm r}^{e{\rm D}}}{m_e}\right)^3 r_d^2 - \left(\frac{m_{\rm r}^{e{\rm H}}}{m_e}\right)^3 r_p^2\right], 
\end{align}
where $h$ is Planck's constant, $\alpha$ is the fine structure constant and $\lambdabar_{\rm C}$ is the reduced Compton wavelength. 
Equating $\Delta\nu^{\rm SM} + \Delta\nu^{\rm NP}$ to $\Delta\nu^{\rm exp}$
and rearranging yields the equation $\Delta = 0$,
where
\begin{align}
\Delta &= 5233.27(42)~\mbox{kHz} + \Delta\nu^{\rm NP} \nonumber \\
&\qquad -\frac{7\alpha^4 m_e c^2}{12 h\lambdabar_{\rm C}^2} \left[\left(\frac{m_{\rm r}^{e{\rm D}}}{m_e}\right)^3 r_d^2 - \left(\frac{m_{\rm r}^{e{\rm H}}}{m_e}\right)^3 r_p^2\right].
 \label{eq:boundC}
\end{align}
For any given values of $g_p$, $g_d$, $g_e$ and $g_\mu$, $\Delta$ is known only within a certain error, $\sigma$, arising primarily from the experimental and theoretical errors on $\Delta \nu^{\rm exp}$, $\Delta \nu^{\rm NP}$,
$r_p$ and $r_d$. 
For given values of the ratio $g_d/g_p$, of the ratio $g_\mu/g_e$ and of the mass $m_{X_0}$, setting
\begin{equation} 
|\Delta| = 1.96\,\sigma
\label{eq:C5}
\end{equation}
then yields the value of $g_eg_p$ beyond which the possibility of an NP interaction is excluded at the 95\% confidence level.

The bounds generated in this way are in close agreement with the ``isotope shift'' results of Figs.~\ref{fig:highmasses} and \ref{fig:fig1_4boxes}. The respective bounds on $g_eg_p$ differ by at most 3\% in the mass range spanned by these figures.

It should be noted that this approach is useful only for cases where $g_d/g_p \not= 1$, 
as for $g_d = g_p$ the New Physics correction term $\Delta\nu^{\rm NP}$ is proportional to $g_eg_p \times (m_{\rm r}^{e{\rm D}} -  m_{\rm r}^{e{\rm H}})/m_e$ rather than to $g_eg_p$. For $r=1$, Eq.~(\ref{eq:C5}) yields bounds on $g_eg_p$ well above the values excluded, e.g., by the results of Fig.~\ref{fig:fig1_10boxes1p0}(e) and (f). 

\section{Further results}
\label{sec:furtherresults}

\subsection{World data with reduced error magnification}
\label{sec:reducederror}

\begin{figure}[t!]
\centering
\includegraphics[width=\columnwidth]{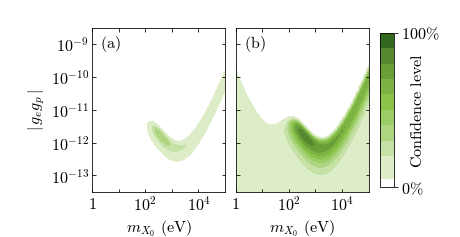}
\caption{
Confidence level that a repulsive NP interaction is compatible with the World spectroscopic data for electronic hydrogen, electronic deuterium, muonic hydrogen and muonic deuterium, assuming that $g_d = g_p$ and that $g_\mu = g_e$. (a): Results obtained without magnification of the experimental uncertainties. (b): Results obtained when the experimental uncertainties are increased by 20\%. As in Fig.~\ref{fig:fig1_10boxes1p0}, the possibility of an NP interaction with parameters falling in a white region is excluded at the 95\% confidence level.
}
\label{fig:smallmag}
\end{figure}
Distributions of confidence levels obtained without or with an error magnification of 20\% are shown in Fig.~\ref{fig:smallmag}, for comparison with Fig.~\ref{fig:fig1_6boxes1p6}(b) (for which the error magnification is 60\%). Fig.~\ref{fig:smallmag}(a) and Fig.~\ref{fig:smallmag}(b) both refer to a repulsive interaction with $r=1$, like Fig.~\ref{fig:fig1_6boxes1p6}(b).

\subsection{Other muonic deuterium models}
\label{sec:othermodels}
\begin{figure}[t!]
\centering
\includegraphics[width=\columnwidth]{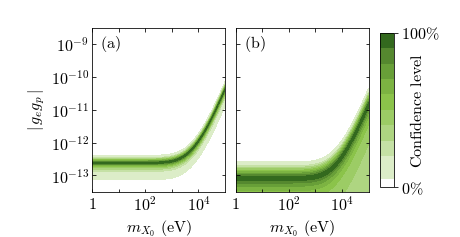}
\caption{
Confidence level that a repulsive NP interaction is compatible with the 1s$_{1/2}$~--~2s$_{1/2}$ transition frequency measured in electronic hydrogen and electronic deuterium and with the nuclear charge radii derived from measurements in muonic hydrogen and muonic deuterium, assuming that $g_d = 2g_p$ and $g_\mu = g_e$. (a): Results obtained when the muonic deuterium structure is described as per Ref.~\cite{Pohl2016}. (b): Results obtained when the muonic deuterium structure is described as per Ref.~\cite{Kalinowski2019}. As in Fig.~\ref{fig:fig1_10boxes1p0}, the possibility of an NP interaction with parameters falling in a white region is excluded at the 95\% confidence level.
}
\label{fig:othermodels}
\end{figure}
Fig.~\ref{fig:othermodels} shows the results obtained for a repulsive NP interaction with $r=2$ when muonic deuterium is described as per Ref.~\cite{Pohl2016} or Ref.~\cite{Kalinowski2019}, rather than as per the more recent theory of Ref.~\cite{Lensky2022} as we do in Fig.~\ref{fig:fig1_4boxes}. Comparing these results to Fig.~\ref{fig:fig1_4boxes}(b) illustrates the sensitivity of these confidence levels on the details of small QED corrections. In particular,  Fig.~\ref{fig:othermodels}(a) excludes the possibility of an attractive NP interaction and strongly points towards the existence of a repulsive NP interaction, whereas Figs.~\ref{fig:othermodels}(b) and \ref{fig:fig1_4boxes}(b) don't.

\mbox{}

\section{Comment on the sensitivity of the isotope shift method in the low mass limit}
\label{sec:limit}
The right-hand side of Eq.~(\ref{eq:deltagegp}) reduces to a simpler form in the low mass limit as the denominator can be worked out in full generality when $m_{X_0} = 0$. Indeed, the NP interaction potential $V_{\rm NP}$ tends to a $1/r$ potential in that limit, which makes it possible to use the virial theorem to calculate the NP shifts $\delta E_{nl}^{\rm NP}$. The result reads
\begin{equation}
\Delta\tilde{\nu}^{\rm NP}_{ba} \approx \frac{8\pi\epsilon_0}{e^2} \, B
{\cal R}\left[r\,{m_{\rm r}^{e{\rm D}} \over m_e} -
{m_{\rm r}^{e{\rm H}}\over m_e}\right]\left({1 \over n_a^2}-{1\over n_b^2}\right).
\end{equation}
Thus $\Delta\tilde{\nu}^{\rm NP}_{ba}$ has the same dependence in $n_a$ and $n_b$ as $\sigma_{ba}^{\rm g}$, in this limit.
Moreover, the numerator reduces to
\begin{displaymath}
1.96\, \sqrt{(\sigma_{ba}^{\rm exp})^2 + 
(\sigma_{ba}^{\rm g})^2}
\end{displaymath}
in the case of transitions between non-s states, since $\sigma_{ba}^{\rm ns} = \sigma_{ba}^{\rm oc} = 0$ for such transitions. This quantity cannot be lower than $1.96\, \sigma_{ba}^{\rm g}$.
The upshot is that for low values of $m_{X_0}$, $\sigma(|g_eg_p|)$ cannot be lower than
$$
\frac{1.96 \times \mbox{0.10~kHz}}{(8\pi\epsilon_0 B {\cal R}/e^2)[r m_{\rm r}^{e{\rm D}}/m_e - m_{\rm r}^{e{\rm H}}/m_e]}$$
for transitions between d-states,
which for $r = 2$ is $2.8\times 10^{-15}$ (a value set by the current uncertainty on the proton and the deuteron masses).


\begin{thebibliography}{99}
\bibitem{shortnote}
There is no need to consider the error on $\Delta\tilde{\nu}^{\rm NP}_{ba}$ here, which arises from the uncertainty on the reduced masses, the use of non-relativistic wave functions, etc., as this error is dwarfed by $\sigma_{{ba}}$.
\end{thebibliography}

\end{document}